\documentstyle[prd,aps,epsfig,preprint,tighten]{revtex}
\newcommand{\tU}{\widetilde{U}}
\newcommand{\tk}{\tilde{k}}
\newcommand{\tK}{\widetilde{K}}
\begin{document}
\draft
\title{		Quasi-energy-independent solar neutrino transitions}
\author{	G.L.\ Fogli,
		E.\ Lisi, and 
		A.\ Palazzo}
\address{     	Dipartimento di Fisica and Sezione INFN di Bari,\\
               	Via Amendola 173, I-70126 Bari, Italy}
\maketitle
\begin{abstract}
Current solar, atmospheric, and reactor neutrino data still allow  oscillation
scenarios where the squared mass differences are all close  to $10^{-3}$
eV$^2$, rather than being hierarchically separated.  For solar neutrinos, this
situation (realized in the upper part of the so-called large-mixing angle
solution) implies adiabatic transitions  which depend weakly on the neutrino
energy and on  the matter density, as well as on the ``atmospheric'' squared
mass difference. In such a regime of ``quasi-energy-independent'' (QEI)
transitions, intermediate between the more familiar
``Mikheyev-Smirnov-Wolfenstein'' (MSW) and energy-independent (EI) regimes, we
first perform analytical calculations of the  solar $\nu_e$ survival
probability at first order in the matter density, beyond the usual hierarchical
approximations.  We then provide accurate, generalized expressions  for the
solar neutrino mixing angles in matter, which reduce to those valid in the MSW,
QEI and EI regimes in appropriate limits. Finally, a representative QEI
scenario is discussed in some detail.
\end{abstract}
\pacs{\\ PACS number(s): 26.65.+t, 14.60.Pq}

\section{Introduction}

The evidence in favor of neutrino mass and mixing \cite{Po67} coming from the
atmospheric anomaly  \cite{SKat,SK00,MACR,Soud} and from the solar neutrino
deficit \cite{NuAs,Cl98,Fu96,Ab99,Ha99,Su00,GNOC,SNOP},  as well as the constraints
on $\nu_e$ mixing from recent reactor searches \cite{CHOO,Palo},  are being
actively investigated both experimentally and theoretically. From the
theoretical point of view, in the absence of new neutrino interactions or new
(sterile) states, the mass-mixing parameters are characterized by the (unitary)
mixing matrix $U$ between the flavor states $\nu_\alpha$  ($\alpha=1,2,3$) and
the mass states $\nu_i$ ($i=1,2,3$),
\begin{equation}
\nu_\alpha = \sum_i U_{\alpha i}\,\nu_i\ ,
\label{Ualphai}
\end{equation}
and by two independent squared mass differences, that we define as
\begin{eqnarray}
\delta m^2 &=& m^2_2-m^2_1\ ,
\label{dm2}\\
m^2 &=& m^2_3-\frac{m^2_1+m^2_2}{2}\ ,
\label{m2}
\end{eqnarray}
as graphically shown in Fig.~1. The two possible independent spectra in Fig.~1
are formally distinguished by the sign of $m^2$, while the sign of $\delta m^2$
can always be taken positive without loss of generality (see also Sec.~III).

From the experimental point of view, combined solar and reactor data analyses
\cite{Pena,Ve01,Mo01} imply an upper limit on the smallest squared mass gap 
$\delta m^2$,
\begin{equation}
\text{solar+reactor}\;\nu:\; 
\delta m^2\lesssim 0.7\times 10^{-3}\text{\ eV}^2\ .
\label{solar}
\end{equation}
Conversely, atmospheric data analyses \cite{SK00,Li00,Pena} imply a  lower
limit on the remaining (larger) squared mass gap $m^2$,
\begin{equation}
\text{atmospheric}\;\nu:\; |m^2|\gtrsim 1.5\times
10^{-3}\text{\ eV}^2\ .
\label{atmos}
\end{equation}
Equations~(\ref{solar}) and (\ref{atmos}) thus imply the  current
phenomenological  limit
\begin{equation}
{\delta m^2}\lesssim \frac{1}{2}{|m^2|} \ .
\label{ha}
\end{equation}

The above constraint is largely fulfilled in several ``hierarchical'' 
oscillation scenarios, where
\begin{equation}
\text{hierarchical cases}\;\longleftrightarrow\;\delta m^2 \ll |m^2|\ ,
\label{hier}
\end{equation}
and the zeroth order approximation in the ratio  $\delta m^2/|m^2|$ is usually
very accurate for both solar \cite{OMso} and laboratory \cite{OMla}
neutrino calculations. At present, however, one cannot exclude that the limit
in Eq.~(\ref{ha}) is saturated, namely, that $\delta m^2$ and $|m^2|$ differ by
less than an order of magnitude. This case, recently considered in  \cite{Stru}
for atmospheric neutrinos, is of great interest for laboratory oscillations
searches (e.g, at future neutrino factories),  where it might lead to
detectable CP-violating effects.

Concerning solar neutrinos, the situation characterized by both $\delta m^2$
and $m^2$ approaching $10^{-3}$ eV$^2$ (from below and from above,
respectively) can affect the upper part of the so-called large mixing angle
(LMA) solution of the solar neutrino problem \cite{OMso,Pena,Kras}. Such a
situation, being relatively close to the regime of ``energy-independent'' 
(EI) 
transitions in vacuum \cite{Grib} 
(established for both  $\delta m^2$ and $|m^2|$
hypothetically above  $\sim 10^{-3}$ eV$^2$) can be called of 
``quasi-energy-independent'' (QEI) transitions
\begin{equation}
\text{QEI} \;\longleftrightarrow\;
\delta m^2 \sim |m^2|\sim O(10^{-3}) \text{\ eV}^2\ ,
\label{qao}
\end{equation}
and is characterized by a mild dependence on  matter effects (``low density''
regime) and on neutrino energy.  Matter effects become increasingly larger for
lower values of $\delta m^2$ ($|m^2|$ being fixed by atmospheric neutrino
data), in  the familiar regime of Mikheyev-Smirnov-Wolfenstein (MSW) 
transitions \cite{MSWo}, where the hierarchical approximation (\ref{hier}) can
be applied.%
\footnote{For even lower values of $\delta m^2$, matter effects play again a
subdominant role (quasivacuum oscillation regime  \protect\cite{QVOr}), but
for  opposite reasons (relatively high matter density).}

At the relatively high values of $\delta m^2$ and $m^2$ implied by  the QEI
regime [Eq.~(\ref{qao})],  oscillation phases are large  and unobservable,
flavor transitions are adiabatic,  and the calculation of the $\nu_e$ survival
probability is  reduced to the calculation of the mixing matrix elements in
matter $\tU_{ei}$.  Although some analytical approximations for the
$\tU_{ei}$'s at low density have been studied in early works on three-flavor
oscillations%
\footnote{An exact analytical diagonalization of the three-flavor neutrino
Hamiltonian in matter is also possible \protect\cite{Ba80,Zagl} but,
unfortunately, the results are not particularly transparent.}
(see, e.g., \cite{Ba80,KuPa,Petc} and refs.\ therein), we think it useful to
revisit  and complete such studies,  especially in order to remove restrictive
hypotheses that have often been used  (e.g., the assumption of sizable
hierarchy $\delta m^2<m^2$ \cite{KuPa} or of small mixing angles \cite{Petc}).

Our paper is structured as follows: In Sec~II we derive analytical expressions
for the $\tU_{ei}$'s in the QEI regime at first order in the matter  density,
with no restrictive assumptions about the neutrino mass hierarchy or mixing, 
and without using a specific parametrization. In Sec.~III we show how to embed
such results in generalized expressions for the mixing angles in matter, which
smoothly connect the familiar MSW regime (for $\delta m^2\ll |m^2|$)  and the
QEI regime (where $\delta m^2 \sim |m^2|$), up to the EI regime.  Such
expressions (written in standard parametrization) may be used to improve the
calculation of the solar neutrino oscillation probability in the high-$\delta
m^2$ fraction of the  LMA solution. Finally, we discuss in Sec.~IV a specific
QEI scenario compatible with present reactor data, and draw our conclusions in
Sec.V.

\section{Parametrization-independent calculations}
\label{S2}

Oscillations in matter are affected by the $\nu_e$ interaction potential $v$ at
the position $x$,
\begin{equation}
v(x)=\sqrt{2}\,G_F\,N_e(x)\ ,
\label{v}
\end{equation}
where $N_e$ is the local electron density \cite{MSWo}. Matter effects are
strong (MSW regime)  when $v$ is of the order of (at least) one of the
wavenumber differences $|k_i-k_j|$, where
\begin{equation}
k_i = \frac{m^2_i}{2E}\ ,
\label{ki}
\end{equation}
$E$ being the neutrino energy. In the QEI regime for solar neutrinos
[Eq.~(\ref{qao})], for  typical neutrino energies, the ratio $v(x)/|k_i-k_j|$
is instead (often much) smaller than unity for $x$ even in the solar core
(regime of ``low density''). Moreover, variations of $v$ along one oscillation
wavelength are extremely small, and the three-family $\nu_e$ survival
probability takes the adiabatic form
\begin{equation}
P_{ee}^{3\nu} = \tU^2_{e1}U^2_{e1}+\tU^2_{e2}U^2_{e2}+\tU^2_{e3}U^2_{e3}\ 
\label{Padiab}
\end{equation}
(see \cite{KuPa} and refs.\ therein), where the  $\tU_{ei}$'s represent the
mixing matrix elements at the production point,%
\footnote{As far as the QEI regime is concerned, the (very low) Earth matter
density at the detection point can be neglected (vacuum approximation).}
and we have taken the three $\nu_e$ mixing matrix elements as real.%
\footnote{This convention does not imply a loss of generality. For complex $U$,
the only difference in our QEI results would be the replacement of $U^2_{ei}$
with $|U_{ei}|^2$ (and analogously for $\tU$). However, one can always choose a
parametrization in which the $|U_{ei}|^2$ do not depend explicitly on the CP
violating phase.  In order to avoid unnecessary book-keeping of $U^*$ terms in
the text, we prefer then to take $U$ real from the beginning.}

The goal of this Section is to calculate the elements $\tU_{ei}$ at first order
in the small parameters $v/|k_i-k_j|$. In Sec.~IIA we discuss in some detail
the spectral decomposition of the Hamiltonian (only rarely used
\cite{Ba80,Aqui}  in the neutrino literature), and in Sec.~IIB we apply it to
the calculation of the $\tU_{ei}$'s  at first order in the matter potential.%
\footnote{ We use the more compact notation $O(v^n)$ as a substitute of
$O(v^n/|k_i-k_j|^n)$.}
No specific parametrization for   $\nu$ masses and mixing is used in Sections
IIA and IIB.

\subsection{Spectral decomposition of the Hamiltonian}

The neutrino Hamiltonian $H$ in the flavor basis $(\nu_e,\nu_\mu,\nu_\tau)$ can
be defined as
\begin{eqnarray}
H &=& UKU^T+V\ ,
\label{H}\\
&=& \tU \tK \tU^T\ ,
\label{tU}
\end{eqnarray}
where $U$ ($\tU$) is the mixing matrix in vacuum (matter), and 
\begin{eqnarray}
V&=& \text{diag}(v,\,0,\,0)\ ,
\label{V}\\
K&=& \text{diag}(k_1,\,k_2,\,k_3)\ ,
\label{K}\\
\tK&=&\text{diag}(\tk_1,\,\tk_2,\,\tk_3)\ ,
\label{tK}
\end{eqnarray}
where the $k_i$'s are given in Eq.~(\ref{ki}), while  the neutrino wavenumbers
in matter $\tilde{k}_i$ represent the  eigenvalues of $H$.

In flavor components, Eq.~(\ref{tU}) reads
\begin{equation}
H_{\alpha\beta} = \sum_i \tk_i\,\tU_{\alpha i}\,\tU_{\beta i}\ ,
\label{HU}
\end{equation}
and thus the products $\tU_{\alpha i}\,\tU_{\beta i}$ can  be identified with
the matrix elements of the projector operators $Q^i$ acting on the
one-dimensional space spanned by the $i$-th eigenvector,
\begin{equation}
 Q^i_{\alpha\beta} = \tU_{\alpha i}\,\tU_{\beta i}\ ,
\label{HP}
\end{equation}
which admit the following  factorization (also called spectral decomposition
theorem of linear algebra):
\begin{equation}
Q^i_{\alpha\beta} =\prod_{j\neq i}
\frac{\tk_j\delta_{\alpha\beta}-H_{\alpha\beta}}{\tk_j-\tk_i}\ .
\label{Q}
\end{equation}

After algebraic manipulations (omitted), which make use of the following two
invariants of the $H$ matrix%
\footnote{The third (unused) invariant is $\tk_1\tk_2\tk_3=\det H$.},
\begin{eqnarray}
\tk_1+\tk_2+\tk_3&=&H_{ee}+H_{\mu\mu}+H_{\tau\tau}\ ,
\label{trH}\\
\tk_1\tk_2 + \tk_2\tk_3 + \tk_3\tk_1 &=& 
H_{ee}H_{\mu\mu}+H_{\mu\mu}H_{\tau\tau}+H_{\tau\tau}H_{ee}\nonumber\\
&&-(H_{e\mu}^2+H_{\mu\tau}^2+H_{\tau e}^2)\ ,
\label{kikj}
\end{eqnarray}
Eq.~(\ref{Q}) can be cast in the form \cite{Aqui}
\begin{equation}
Q^i_{\alpha\alpha}=\frac
{(\tk_i-H_{\beta\beta})(\tk_i-H_{\gamma\gamma})-H^2_{\beta\gamma}}
{(\tk_i-\tk_j)(\tk_i-\tk_n)}
\label{Qaa}
\end{equation}
for the diagonal elements, and to
\begin{equation}
Q^i_{\alpha\beta}=\frac
{H_{\alpha\beta}(\tk_i-H_{\gamma\gamma})+H_{\alpha\gamma}H_{\beta\gamma}}
{(\tk_i-\tk_j)(\tk_i-\tk_n)}
\label{Qab}
\end{equation}
for the off-diagonal elements, where $(\alpha,\,\beta,\,\gamma)$ are
permutations of $(e,\,\mu,\,\tau)$ and $(i,\,j,\,n)$ are permutations of
$(1,\,2,\,3)$.%
\footnote{Notice that, in the first of Ref.~\protect\cite{Aqui}, there is a
sign misprint in the expression of $Q^i_{\alpha\beta}$.}
By comparing Eq.~(\ref{HP}) with  Eq.~(\ref{Qaa}), one finally gets an explicit
expression for the $\tU^2_{ei}$'s  as a function of the eigenvalues of $H$ and
of its ($\mu,\tau$) submatrix elements,
\begin{equation}
\tU^2_{ei} = \frac
{(\tk_i-H_{\mu\mu})(\tk_i-H_{\tau\tau})-H^2_{\mu\tau}}
{(\tk_i-\tk_j)(\tk_i-\tk_n)}\ .
\label{UU}
\end{equation}

\subsection{Expressions for $\tU$ valid at first order in $v$}

At first order in $v$, the eigenvalues $\tilde{k}_i$ of $H$ are most easily
calculated in the vacuum mass basis $\nu_i$, where the Hamiltonian is
\begin{equation}
H'=K+U^T VU\ ,
\label{H'}
\end{equation}
and the eigenvalue equation [$\det(H'-\tk I)=0$]
turns out to be already factorized,
\begin{equation}
\prod_i (k_i-\tk_i+vU^2_{ei}) +O(v^2) =0\ ,
\label{eigen}
\end{equation}
leading to the (known, see \cite{Ba80}) result
\begin{equation}
\tk_i = k_i + vU^2_{ei} + O(v^2)\ .
\label{eigenk}
\end{equation}

By means of Eq.~(\ref{eigenk}) and Eq.(\ref{UU}) we finally obtain, after
somewhat lengthy but straightforward algebra,
\begin{equation}
\tU^2_{ei} = U^2_{ei}\left[ 
1+2v\left(
\frac{U^2_{ej}}{k_i-k_j} + \frac{U^2_{en}}{k_i-k_n} 
\right)
\right] + O(v^2)\ ,
\label{tUi}
\end{equation}
where $(i,j,n)$ are permutations of $(1,2,3)$. This is our basic result in the
QEI regime.

Notice that Eq.~(\ref{tUi}) depends explicitly on the {\em squared mass
differences\/} $m^2_i-m^2_j$,  as it should, without restrictive (i.e.,
hierarchical) assumptions about their relative  magnitude.  Notice also that
Eq.~(\ref{tUi}) holds for generic values of the elements $U^2_{ei}$ (within the
unitarity constraint $\sum_i U^2_{ei}=1$).%
\footnote{ By making the further assumption of small $U^2_{e2}$ and $U^2_{e3}$,
and by applying the parametrization adopted in \protect\cite{Petc} for the
matrix $U$, Eq.~(\protect\ref{tUi}) reproduces the results found at first order
in $(v,U^2_{e2},U^2_{e3})$ in \protect\cite{Petc} for the low-density case.}

\section{QEI Results in standard parametrization}

In the standard notation  for  the neutrino mixing matrix $U$ \cite{KuPa},
\begin{equation}
U=U(\theta_{12},\,\theta_{13},\,\theta_{23})=U(\omega,\,\varphi,\,\psi)\ .
\label{Uparam}
\end{equation}
the mixing matrix elements relevant to solar neutrinos read 
\begin{eqnarray}
U^2_{e1} &=& \cos^2\varphi\cos^2\omega\ ,\\
U^2_{e2} &=& \cos^2\varphi\sin^2\omega\ ,\\
U^2_{e3} &=& \sin^2\varphi\ .
\label{U2}
\end{eqnarray}
and thus depend only the angles  $\varphi$ and $\omega$.%
\footnote{ The angle $\psi$ corresponds to a rotation in the
$(\nu_\mu,\nu_\tau)$ subspace, which is unobservable in solar neutrino
experiments. The possible CP violating phase can also be put in such subspace
and rotated away, as far as solar $\nu$'s are concerned. This justifies the
choice of real $U$ for solar neutrinos.}

By using an analogous notation for the mixing matrix and angles in matter
(denoted by a tilde) one has
\begin{eqnarray}
\tan^2\tilde{\omega} &=& \frac{\tU^2_{e2}}{\tU^2_{e1}}\ ,
\label{t2omega}\\
\sin^2\tilde{\varphi} &=& \tU^2_{e3}\ ,
\label{s2phi}
\end{eqnarray}
and Eq.~(\ref{Padiab}) for the QEI probability reads
\begin{equation}
P_{ee}^{3\nu} = \cos^2\tilde{\varphi}\cos^2
\varphi(\cos^2\tilde{\omega}\cos^2\omega+
\sin^2\tilde{\omega}\sin^2\omega)
+\sin^2\tilde{\varphi}\sin^2\varphi\ .
\label{Peegen}
\end{equation}

In order to study the symmetry properties of  $P_{ee}^{3\nu}$, we introduce 
three (commuting) transformations $T$, 
\begin{eqnarray}
T_\omega \;&\Longleftrightarrow&\; \omega \to \pi/2 - \omega\ ,
\label{Tomega}\\
T_{\delta m^2} \;&\Longleftrightarrow&\; \delta m^2 \to -\delta m^2\ ,
\label{Tdeltam2}\\
T_{ m^2} \;&\Longleftrightarrow&\; m^2 \to -m^2\ ,
\label{Tm2} 
\end{eqnarray}
and also  define two independent neutrino wavenumbers [associated to the
squared mass gaps in Eqs.~(\ref{dm2},\ref{m2})]
\begin{eqnarray}
\delta k &=&\frac{\delta m^2}{2E}\ ,
\label{deltak}\\
k &=& \frac{m^2}{2E}
\label{k}\ .
\end{eqnarray}

In terms of the previous notation, Eqs.~(\ref{tUi}) and (\ref{t2omega})  imply
the following $O(v)$ expression for $\tan^2 \tilde{\omega}$
\begin{equation}
\tan^2\tilde{\omega}
= \tan^2\omega \left(1+ 2\frac{v\cos^2\varphi}{\delta k}f_\varphi
\right)+O(v^2)\ ,
\label{tomega}
\end{equation}
where 
\begin{equation}
f_\varphi =1-
\frac{\displaystyle\left(
\frac{\delta k}{k}\right)^2\tan^2\varphi}{1-\left(\displaystyle
\frac{\delta k}{2 k}\right)^2}
\label{fphi}
\end{equation}
The above two equations are invariant under the combined symmetry operation
$T_{\delta m^2} T_\omega $, which simply means that $\delta m^2$ can always be
taken positive, as far as $\omega$ is taken in its full range $[0,\pi/2]$. The
expression for $\tilde\omega$ is also invariant under the operation $T_{m^2}$,
which means that $\tilde{\omega}$ carries no information about the difference
between the two spectra in Fig.~1. Such information is carried instead by
$\tilde{\varphi}$.

In fact, the angle $\tilde{\varphi}$ can be expressed, through 
Eqs.~(\ref{tUi}) and (\ref{s2phi}), as
\begin{equation}
\sin^2\tilde{\varphi}
= \sin^2\varphi \left(1+ 2\frac{v\cos^2\varphi}{k}g_\omega
\right)+O(v^2)\ ,
\label{tphi2}
\end{equation}
where
\begin{equation}
g_\omega=
\frac{1-\displaystyle
\frac{\delta k}{2k}\cos2\omega}{1-\left(\displaystyle
\frac{\delta k}{2k}\right)^2}\ .
\label{fomega}
\end{equation}

The above two equations are symmetric under $T_{\delta m^2}T_\omega$, but {\em
not\/} under $T_{m^2}$. Therefore, if $\varphi$ is nonzero and if oscillations
take place in the QEI regime, the two spectra in Fig.~1 can be
distinguished---in principle---by solar neutrino data. In practice, the
uncertainties affecting the solar neutrino phenomenology  currently prevent
such discrimination \cite{Ve01,Prog}.  However, the sensitivity to $m^2$  (and
to its sign) might be enhanced in the near future if  the solar neutrino
parameters $(\delta m^2,\omega)$ were confirmed (and accurately measured) in
the LMA region by reactor oscillation searches \cite{Barg}.

Notice that the previous expressions for the mixing angles allow to rewrite
Eq.~(\ref{Peegen}) in the form 
\begin{equation}
P_{ee}^{3\nu}= \cos^2\tilde{\varphi}\cos^2\varphi
\left(P_{ee}^{2\nu}\right)_{v\to v f_\varphi\cos^2\varphi }
+\sin^2\tilde{\varphi}\sin^2\varphi\ +O(v^2),
\label{Pee32}
\end{equation}
where $P^{2\nu}_{ee}$ represents  the adiabatic survival probability for the
two-flavor subcase (namely, $\varphi=0$ and 
$P^{2\nu}_{ee}=\cos^2\tilde{\omega}\cos^2\omega +
\sin^2\tilde{\omega}\sin^2\omega$), provided that the effective electron
density is taken as $f_\varphi \cos^2\varphi N_e$ in the calculation of
$\tilde{\omega}$ [see Eq.~(\ref{tomega})].  The above recipe for the QEI
probability in three families is formally equivalent the one applicable in the
MSW regime \cite{Ptre,KuPa,Shis,OMso},  modulo the additional factor
$f_\varphi$ multiplying the effective density.

\section{Generalized expressions for mixing angles in matter}

In this section we provide accurate expressions for the mixing angles in
matter, generalizing those valid in the MSW, QEI, and EI regimes,
which are recovered as specific subcases. Therefore, such expressions
can be particularly useful for solar neutrino calculations spanning the three
aforementioned regimes. 

Let
us express the QEI results [Eqs.~(\ref{tomega}) and (\ref{tphi2})]  in terms of
the  variables $\sin^2 2\tilde{\omega}$ and $\sin^2 2\tilde{\varphi}$,
\begin{eqnarray}
\sin^2 2\tilde{\omega} (\text{\small QEI}) &\simeq & \sin^2 2\omega
\left(1+2\cos2\omega\,\frac{v\cos^2\varphi}{\delta k}f_\varphi\right) 
\label{sin2omega}\ ,\\
\sin^2 2\tilde{\varphi}(\text{\small QEI}) &\simeq & \sin^2 2\varphi
\left(1+2\cos2\varphi\,\frac{v}{k}g_\omega\right)
\label{sin2phi}\ ,
\end{eqnarray}
and write again
the three-flavor QEI probability [Eq.~(\ref{Pee32})],
\begin{equation}
P_{ee}^{3\nu} (\text{\small QEI}) \simeq \cos^2\tilde{\varphi}\cos^2\varphi
\left(P_{ee}^{2\nu}\right)_{v\to v f_\varphi\cos^2\varphi }
+\sin^2\tilde{\varphi}\sin^2\varphi\ .
\label{PeeQAO}
\end{equation}
We remind that Eqs.~(\ref{sin2omega})--(\ref{PeeQAO}) have been derived at
first order in $v/k$ and $v/\delta k$ (implying small matter effects),  with
no (hierarchical) restriction on the relative magnitude of $k$ and $\delta k$.

The EI transition regime is recovered from the QEI case
in the 
limit of very large squared mass
differences, i.e., at zeroth order in $v/k$ and $v/\delta k$
(so that $\sin^2\tilde{\omega}=\sin^2\omega$ and 
$\sin^2\tilde{\varphi}=\sin^2\varphi$).
The EI probability reads then
\begin{eqnarray}
P_{ee}^{3\nu} (\text{\small EI}) &\simeq& \cos^4\varphi
(\cos^4\omega+\sin^4\omega)
+\sin^4\varphi \label{PeeEI}\\
&=& U^4_{e1}+U^4_{e2}+U^4_{e3}\ .
\end{eqnarray}

The MSW regime is instead characterized  by 
$\delta k\sim O(v)$ (strong matter effects). 
In this case, the expressions for
$\tilde{\omega}$, $\tilde{\varphi}$, and $P_{ee}^{3\nu}$
are often derived under a strictly hierarchical hypothesis
(i.e., at zeroth order
in  both $\delta k/k$ and  $v/k$), but with no further restriction
on the value of $v/\delta k$,  giving the well-known results
\begin{eqnarray}
\sin^2 2\tilde{\omega} (\text{\small MSW})&\simeq & \frac{\sin^2 2\omega}
{\left(\displaystyle \cos2\omega-\frac{v\cos^2\varphi}
{\delta k}\right)^2 +
\sin^2 2\omega} 
\label{sin2omega2OM}\ ,\\
\sin^2 2\tilde{\varphi} (\text{\small MSW}) &\simeq & \sin^2 2\varphi\ ,
\label{sin2phi2OM}
\end{eqnarray}
and
\begin{equation}
P_{ee}^{3\nu} (\text{\small MSW}) \simeq \cos^4{\varphi}
\left(P_{ee}^{2\nu}\right)_{v\to v \cos^2\varphi }
+\sin^4{\varphi}\ ,
\label{PeeMSW}
\end{equation}
where $P^{2\nu}_{ee}$ is the probability for the two-family subcase, containing the
so-called crossing probability $P_c$ in the nonadiabatic MSW case   (see
\cite{Ptre,KuPa,Shis,OMso} and references therein).

Less often, the above MSW expressions are improved by including 
the lowest-order effects of nonzero $v/k$ and $\delta k/k$ \cite{Cons},
described by the (primed) expressions \cite{KuPa}
\begin{eqnarray}
\sin^2 2\tilde{\omega} (\text{\small MSW}')&\simeq & \frac{\sin^2 2\omega}
{\left(\displaystyle \cos2\omega-\frac{v\cos^2\varphi}
{\delta k}\right)^2 +
\sin^2 2\omega} 
\label{sin2omega2KP}\ ,\\
\sin^2 2\tilde{\varphi} (\text{\small MSW}')&\simeq & \frac{\sin^2 2\varphi}
{\left(\displaystyle \cos2\varphi-\frac{v}{ k+\frac{\delta k}{2}
\cos2\omega}\right)^2 +
\sin^2 2\varphi} \ .
\label{sin2phi2KP}
\end{eqnarray}
and
\begin{equation}
P_{ee}^{3\nu} (\text{\small MSW}') \simeq  \cos^2\tilde{\varphi}\cos^2\varphi
\left(P_{ee}^{2\nu}\right)_{v\to v \cos^2\varphi }
+\sin^2\tilde{\varphi}\sin^2\varphi\ .
\label{PeeMSW'}
\end{equation}

By comparing all the previous expressions,  derived under different
approximations in the QEI, EI, and MSW(') cases,
we find that they can be considered
as appropriate
subcases of the following generalized expressions for the mixing angles in
matter,
\begin{eqnarray}
\sin^2 2\tilde{\omega}  &\simeq & \frac{\sin^2 2\omega}
{\left(\displaystyle \cos2\omega-\frac{v\cos^2\varphi}
{\delta k}f_\varphi\right)^2 +
\sin^2 2\omega}\ ,\label{sin2omega2}\\
\sin^2 2\tilde{\varphi}  &\simeq & \frac{\sin^2 2\varphi}
{\left(\displaystyle \cos2\varphi-\frac{v}{k}g_\omega\right)^2 +
\sin^2 2\varphi}\ ,
\label{sin2phi2}
\end{eqnarray}
and for  three-family oscillation probability (in terms of the two-family one)
\begin{equation}
P_{ee}^{3\nu}  \simeq \cos^2\tilde{\varphi}\cos^2\varphi
\left(P_{ee}^{2\nu}\right)_{v\to v f_\varphi\cos^2\varphi }
+\sin^2\tilde{\varphi}\sin^2\varphi\ .
\label{Peefinal}
\end{equation}

In particular, using the above generalized expressions
(\ref{sin2omega2})--(\ref{Peefinal}), it turns out that:  $(i)$ The QEI
approximation [Eqs.~(\ref{sin2omega})--(\ref{PeeQAO})] is recovered through a
first-order expansion in $v/k$ and $v/\delta k$; $(ii)$ The EI case
[Eq.~(\ref{PeeEI})] is reproduced in the limit $v/\delta k,\,v/k \to0$; $(iii)$
The usual MSW expressions  [Eqs.~(\ref{sin2omega2OM})--(\ref{PeeMSW})] are
recovered in the (strictly hierarchical) limit $k\to\infty$; and $(iv)$ The
improved MSW' expressions [Eqs.~(\ref{sin2omega2KP})--(\ref{PeeMSW'})] are
recovered through a first-order expansion in $\delta k/k$. Therefore,
Eqs.~(\ref{sin2omega2}), (\ref{sin2phi2}) and (\ref{Peefinal}), together with
definitions in Eqs.~(\ref{fomega}) and (\ref{fphi}), provide useful
generalizations of $P^{3\nu}_{ee}$ and of the mixing angles in matter
$\tilde{\omega}$ and $\tilde{\varphi}$, smoothly interpolating from the
familiar MSW regime (where $P^{3\nu}_{ee}$ depends on $\delta m^2$ only)  to
the QEI regime  (where $P^{3\nu}_{ee}$ depends on both $\delta m^2$ and $m^2$)
and further up to the EI regime (independent on both $\delta m^2$ and $m^2$).

We have also performed the following numerical check: For many representative
$(\delta m^2,\,m^2,\,\omega,\,\varphi,\,E)$ values of phenomenological interest
in the QEI regime, we have computed $P_{ee}^{3\nu}(E)$ both through the
analytical expressions in  Eqs.~(\ref{sin2omega2})--(\ref{Peefinal}) and
through Eq.~(\ref{Padiab}) with $U$ obtained by numerical diagonalization.  We
find differences (often much) smaller than $10^{-3}$ in $P_{ee}^{3\nu}$.

Analogously, for the same previous $(\delta m^2,\,m^2,\,\omega,\,\varphi,\,E)$
values, we have tested the accuracy of the MSW' approximations
[Eqs.~(\ref{sin2omega2KP})--(\ref{PeeMSW'})]  which, being derived under the
hierarchical assumption of small $\delta k/k$, might not properly work in the
QEI case. We typically find only a slight worsening of the accuracy (no more
than a factor of two),  as compared with the previous check.  Such a slight
worsening is maximized for the highest values of $\delta m^2$ and $\sin^2
\varphi$ allowed by current neutrino phenomenology (about $0.7\times 10^{-3}$
eV$^2$ and $0.05$,  respectively, see later). Notice that, for $\varphi\to 0$,
Eqs.~(\ref{sin2omega2KP})--(\ref{PeeMSW'})  and
Eqs.~(\ref{sin2omega2})--(\ref{Peefinal}) tend to the same $2\nu$ limit, where
genuine $m^2$-induced QEI effects disappear.

In conclusion, Eqs.~(\ref{sin2omega2})--(\ref{Peefinal}) provide a general and
accurate prescription to calculate the mixing angles and the $\nu_e$ survival
probability in matter, smoothly  interpolating between the MSW, QEI and EI
regimes. The prescription  is most useful for nonvanishing $\varphi$.
Preliminary applications of this computing recipe in the analysis of the
current solar neutrino phenomenology have been presented in \cite{Ve01,Mo01}
and will be discussed in a separate work  \cite{Prog}. In the next section, we
just focus on a representative QEI case compatible with present reactor bounds.

\section{Discussion of a representative QEI scenario}

In this Section we discuss a representative spectrum of squared mass
differences leading to the QEI case for solar neutrinos, namely
\begin{equation}
(\delta m^2, m^2) = (0.6,\pm 1.5) \times 10^{-3}\text{\ eV}^2\ ,
\label{spectrum}
\end{equation}
where the sign of $m^2$ discriminates the two options in Fig.~1. The above
value for $\delta m^2$ is marginally
allowed in the upper part of the LMA solution to the
solar neutrino problem, while the value of $|m^2|$ is allowed in the lower
range of the oscillation solution to the atmospheric neutrino anomaly
\cite{SK00}.   Concerning neutrino mixing, we choose a small (but nonzero)
value for $\varphi$, and a value for $\omega$ within the LMA solution (as well as
its octant-symmetric  value $\pi/2-\omega$),
\begin{eqnarray}
\tan^2\varphi&=&0.04\ ,\\
\tan^2\omega &=&0.5\  (2.0)\ .
\label{mixing}
\end{eqnarray}

With the above choice for the mass-mixing parameters, it turns that two
different squared mass gaps  $(m^2\pm \delta m^2/2)$ are in the sensitivity
range of reactor experiments such as CHOOZ \cite{CHOO} and Palo Verde
\cite{Palo}  ($\gtrsim 0.7 \times 10^{-3}$ eV$^2$), so that the usual bounds
derived for the $2\nu$ \cite{SK00} case or for $3\nu$ case with $\delta
m^2\simeq 0$  \cite{3atm} are not immediately applicable, and require a
dedicated study.%
\footnote{In such a study it is sufficient to consider only CHOOZ data,
the Palo Verde data being slightly less restrictive on the same
mass-mixing parameters.}

\subsection{CHOOZ constraints}

The general $3\nu$ survival probability for electron antineutrinos at reactors
(in vacuum) reads
\begin{eqnarray}
P_{ee}^{\rm reac} =  1 &-& 4 \cos^4\varphi\sin^2\omega\cos^2\omega
\,\sin^2\left(\frac{\delta k}{2}x\right )\nonumber\\ 
&-& 4 \sin^2\varphi\cos^2\varphi\sin^2\omega
\,\sin^2\left(\frac{k-\delta k/2}{2}x\right )\nonumber\\ 
&-& 4 \sin^2\varphi\cos^2\varphi \cos^2\omega
\,\sin^2\left(\frac{k+\delta k/2}{2}x\right )\ ,
\label{Pchooz}
\end{eqnarray}
where $x$ is the baseline. The above expression is invariant under the symmetry
transformations $T_{\delta m^2}T_{\omega}$  $T_{\delta m^2}T_{m^2}$, and 
$T_{m^2}T_{\omega}$ [defined in  Eqs.~(\ref{Tomega})--(\ref{Tm2})], implying
that the two spectra in Fig.~1 cannot be distinguished by reactor neutrino data
(while they can be by solar $\nu$ data in the QEI regime, at least in
principle).

In \cite{Pena}, Eq.~(\ref{Pchooz}) has been used in global $3\nu$ oscillation
fits by using the {\em total\/} CHOOZ rate \cite{CHOO}. However, since the
low-energy part of the CHOOZ  spectrum is more sensitive  to relatively low
values of $\delta m^2$, we prefer to use the full CHOOZ data set (i.e., the
binned spectra from the two reactors) rather than the total rate only. In
particular, we have accurately reproduced the so-called 
``$\chi^2$ analysis A'' of
\cite{CHOO}, by using two 7-bin positron spectra and one constrained 
normalization parameter, for a total of $14+1-1=14$ independent (but correlated)
data. We obtain very good agreement with Fig.~9 in \cite{CHOO} for
the two-flavor subcase (not shown).%
\footnote{The analysis A in  \protect\cite{CHOO} also introduces a constrained
energy scale shift, that we omit for lack of published information. Its effect
seems to be small {\em a posteriori}, given that our bounds reproduce those of 
\protect\cite{CHOO} anyway.}

Using our binned $\chi^2$ analysis for CHOOZ, and setting $\tan^2\varphi=0.04$,
$\tan^2\omega=0.5$, and $\delta m^2=0.6\times 10^{-3}$ eV$^2$, we obtain 
$\chi^2/N_{\rm DF}=15.5/14$  for $m^2=+1.5\times 10^{-3}$ eV$^2$ and 
$\chi^2/N_{\rm DF}=13.4/14$  for $m^2=-1.5\times 10^{-3}$ eV$^2$. Due to the
symmetry $T_{m^2}T\omega$, the previous $\chi^2$ values also apply for 
$\tan^2\omega=2$ by replacing $\pm m^2$ with $\mp m^2$. In any case, the choice
of parameters adopted in Eqs.~(\ref{spectrum})--(\ref{mixing}) gives
$\chi^2/N_{\rm DF}\simeq 1$,  and thus  passes the goodness-of-fit test.

\subsection{The QEI probability}

Figure~2 shows the solar $\nu_e$ survival probability  derived from
Eqs.~(\ref{sin2omega2})--(\ref{Peefinal}) (and averaged over the $^8$B
production region for definiteness) as a function of neutrino energy.  The QEI
cases in  Eqs.~(\ref{spectrum})--(\ref{mixing}) are represented by either
dot-dashed lines ($m^2>0$) or dashed lines ($m^2<0$). Such lines collapse to a
single (solid) line for $m^2\to\infty$, which provides the usual
($m^2$-independent) hierarchical limit. If one also takes $\delta
m^2\to\infty$, the energy dependence is averaged out (EI regime) and the
probability becomes constant (dotted, horizontal line).

Notice that the sizable QEI deviation from the constant EI case, induced by the
finite value of $\delta m^2$, changes sign from the first to the second octant
of $\omega$. Such octant asymmetry, which asymptotically disappears for
increasing values of $\delta m^2$, is still effective in the upper part of the
LMA solution ($\delta m^2\lesssim 10^{-3}$ eV$^2$), where it becomes manifest 
as a slight local preference of current solar $\nu$ data fits for
$\tan^2\omega<1$. This preference is mainly driven by the low Chlorine rate
\cite{Cl98}, which favors relatively low values of $P_{ee}$ (realized in Fig.~2
by the lines at $\tan^2\omega=1/2$).  The subleading QEI deviation due to the
finite value and sign  of $m^2$  (which splits the solid line into the dashed
and dot-dashed curves in Fig.~2) plays instead a minor role in the current
solar $\nu$ phenomenology \cite{Ve01,Mo01}, which does not show a statistically
significant preference for one of the two spectra in Fig.~1. However, the
situation might be improved in the near future \cite{Prog}, should the $(\delta
m^2,\omega)$ parameters be confirmed and narrowed in the upper part of the LMA
region by  KamLand \cite{Barg,Kaml} and by the second-generation solar neutrino
experiments SNO \cite{SNOP,SNO1} 
and BOREXINO \cite{Bore}. In this case (provided
that $\varphi$ is nonvanishing), residual $m^2$-induced QEI corrections  might
play a role in accurate calculations of $P^{3\nu}_{ee}$.

\section{Conclusions}

We have performed (perturbative)  analytical calculations of the solar $3\nu$
survival probability $P^{3\nu}_{ee}$ in the regime of quasi-energy-independent
(QEI) transitions,  intermediate between the more familiar MSW and 
energy-independent (EI) regimes, 
and characterized by squared mass differences all
close to $\sim 10^{-3}$ eV$^2$. We have generalized well-known MSW expressions
for $P^{3\nu}_{ee}$ and for the mixing angles $\omega=\theta_{12}$  and
$\varphi=\theta_{13}$  (valid for hierarchical mass differences) in a form
which smoothly matches the corresponding expressions for the (nonhierarchical) 
QEI regime. Our main results, summarized in 
Eqs.~(\ref{sin2omega2})--(\ref{Peefinal}) [together with the definitions in
Eqs.~(\ref{fphi}) and (\ref{fomega})], represent an accurate and simple recipe
that can be used to improve current calculations of $P^{3\nu}_{ee}$ in the
upper part of the LMA solution, where the QEI regime is effective and,
 for nonvanishing $\varphi$,
there might be a residual sensitivity of solar neutrino transitions to the
``atmospheric'' squared mass difference.

\section*{Acknowledgments}

We thank D.\ Montanino for useful discussions and suggestions. This work was
supported in part by the Italian INFN and MIUR under the ``Astroparticle
Physics'' project.


\newcommand{\InsertFigure}[2]{\newpage\phantom{.}
\vspace*{0.cm}\begin{center}\mbox{%
\epsfig{bbllx=2truecm,bblly=2truecm,bburx=19.5truecm,bbury=26.5truecm,%
height=20.truecm,figure=#1}}\end{center}\vspace*{-0.8truecm}%
\parbox[t]{\hsize}{\small\baselineskip=0.5truecm\hskip0.5truecm #2}}

\InsertFigure{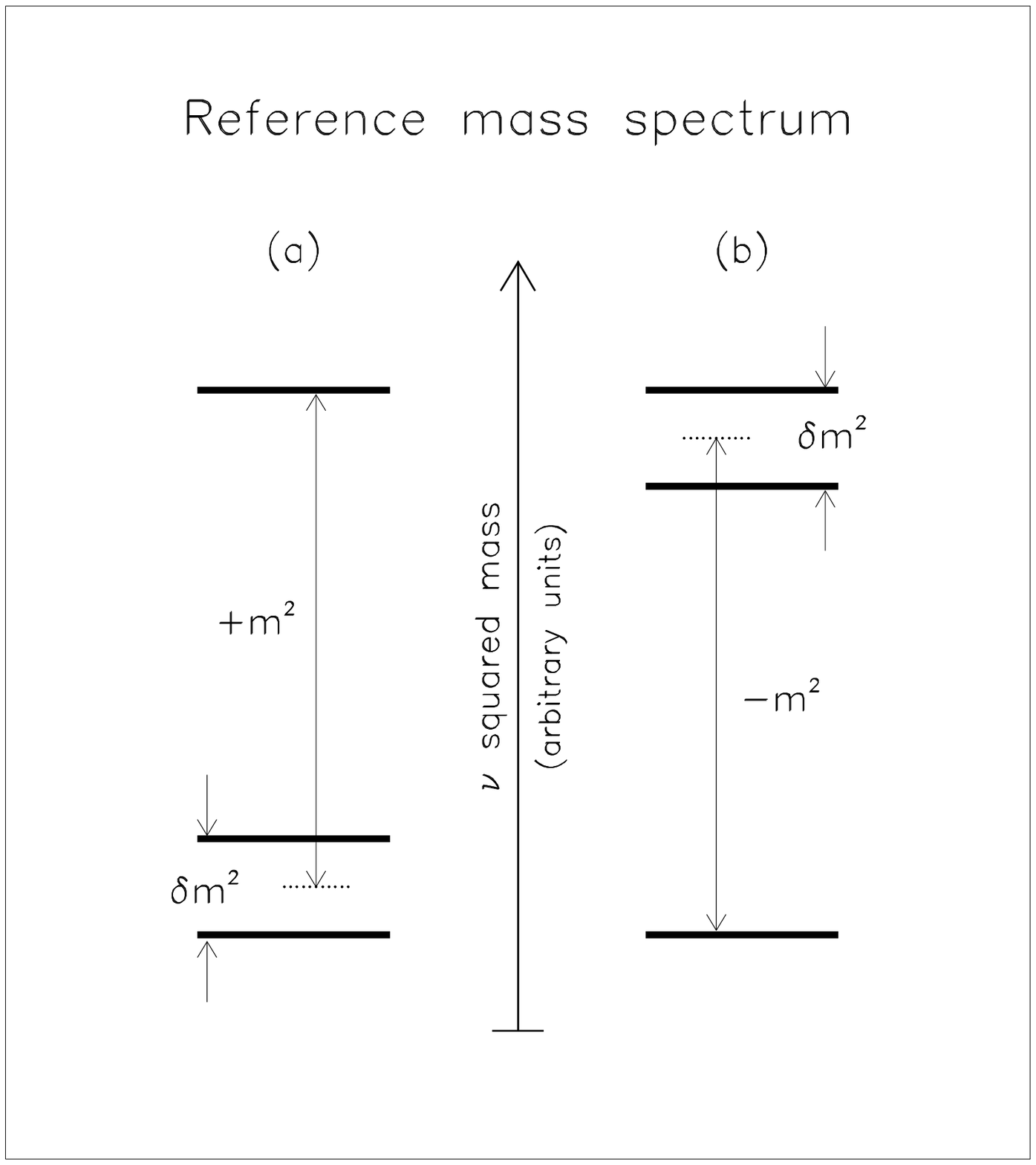}%
{FIG.~1. Neutrino mass spectrum, together with our notation for the squared
mass differences $\delta m^2$ and $m^2$. When both such parameters approach
$10^{-3}$ eV$^2$, solar neutrinos undergo quasi-energy-independent
 oscillations. In the
QEI regime, both $m^2$ and the relative sign between $\delta m^2$ and $m^2$
become observable in matter and can, in principle, discriminate the two
possible options in figure.}
\InsertFigure{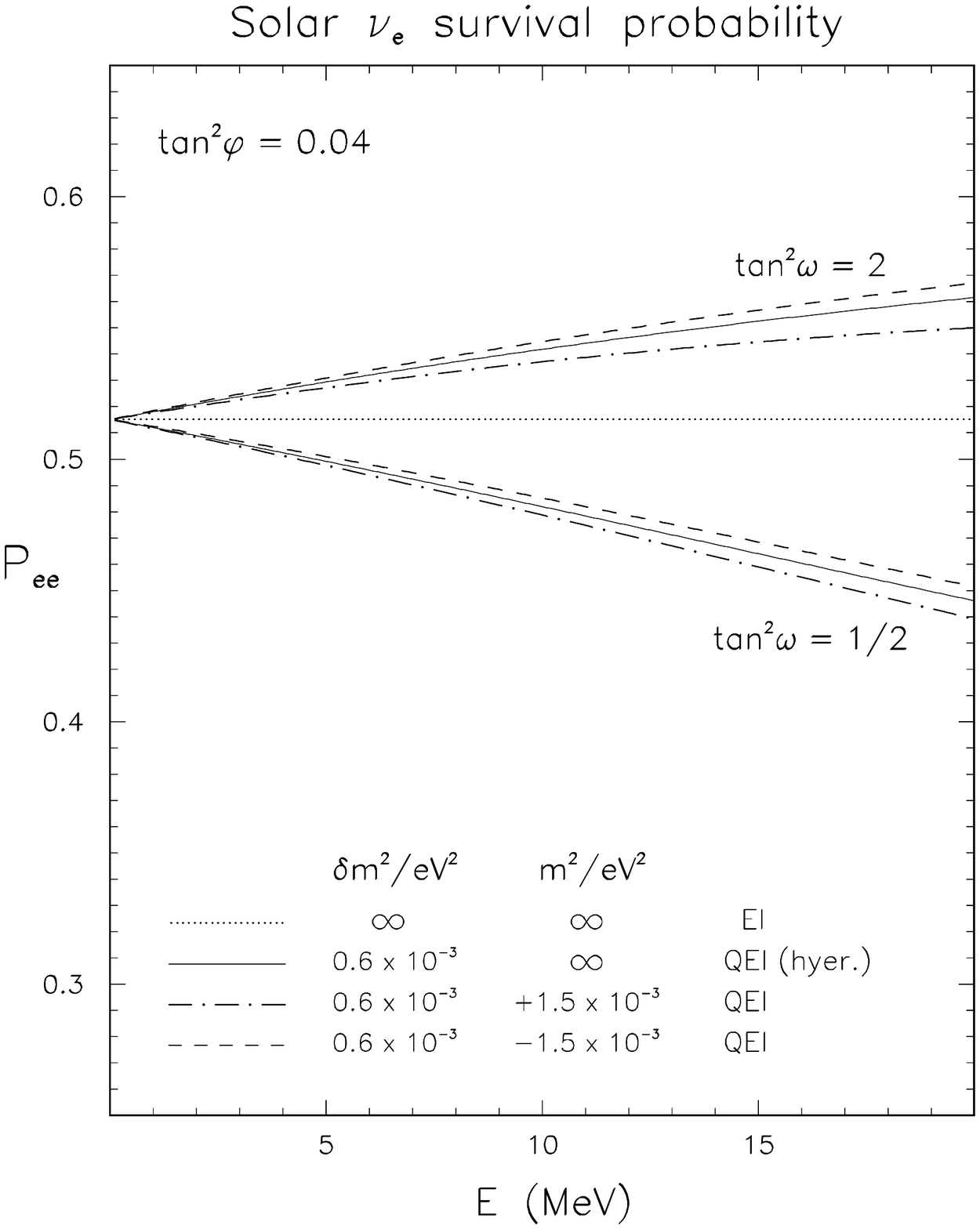}%
{FIG.~2. QEI effects on $P^{3\nu}_{ee}$, averaged over the $^8$B production
region, for $(\delta m^2,\,m^2)=(0.6\,,\pm1.5)\times 10^{-3}$ eV$^2$, together
with the asymptotic behavior for $|m^2|\to\infty$ (hierarchical case) and for
both $\delta m^2$ and $|m^2|$ large (averaged oscillations). The mixing
parameter $\tan^2\varphi$ is set at the value 0.04, while $\tan^2\omega$ is taken
to be either 1/2 or 2 (corresponding to octant-symmetric values of $\omega$).
The QEI cases in this figure are allowed by CHOOZ spectral data.}

\end{document}